\documentclass[aps,pre,preprint,groupedaddress,epsfig,superscriptaddress,showpacs]{revtex4-1}
\usepackage{graphicx}
\usepackage{bm}
\usepackage{CJKutf8}
\usepackage{caption} 
\usepackage{color}
\usepackage{ragged2e}
\usepackage{amsmath}
\renewcommand{\raggedright}{\leftskip=0pt \rightskip=0pt plus 0cm}
\begin{document}

%\preprint{}

\title{Single-orientation Crystalline Domains of Active Brownian Particles Lead to Collective Motions}


%\author{}
%\thanks{}
%\affiliation{}
\author{Cheng Yang \footnote{E-Mail: yangcheng@mtc.edu.cn
}}
\affiliation{School of Physics and Electronic Information, Mianyang Teachers' College, Mianyang 621000, China}

\author{Qiandong Dai}
\affiliation{School of Physics and Electronic Information, Mianyang Teachers' College, Mianyang 621000, China}

\author{Shun Xu \footnote{E-Mail: xushun@sccas.cn
}}
\affiliation{Computer Network Information Center, Chinese Academy of Sciences, Beijing 100083, China}

\author{Xin Zhou \footnote{E-Mail: xzhou@ucas.ac.cn
}}
\affiliation{School of Physical Sciences, University of Chinese Academy of Sciences, Beijing 100049, China.}


\date{\today}

\begin{abstract}
Active Brownian particles, even without attractive and anisotropic inter-particle interactions, can form a high-density phase featuring structure-ordered domains as well as collective motion regions under thermal noise. However, the mechanism, particularly the relationship between the motion and structure, remains unclear. In this study, we show that the motion-correlation regions are spatially coincident with the single-orientation crystalline domains. Each domain translates or rotates as a whole due to the net active force or torque acting upon it, allowing relative motions between these crystalline domains. The particles at domain boundaries usually have the active forces pointing inward, which helps to stabilize these structure-ordered domains and their corresponding collective motion regions.  
\end{abstract}

%\keywords{}

\maketitle

\section{Introduction}
Active matter---both living and non-living---utilizes energy to maintain systematic motion~\citep{bechinger2016active,das2020introduction,marchetti2013hydrodynamics,
ramaswamy2010mechanics}, with examples spanning from biological entities such as microorganisms\citep{angelini2010cell,garcia2015physics,liu2021shannon,Zhang_2009,PhysRevLett102168101} and animals \citep{buhl2006disorder,ballerini2008interaction} to man-made imitations~\citep{buttinoni2013dynamical,palacci2013living,
theurkauff2012dynamic,deseigne2010collective,wang2021emergent,liu2020oscillating,yang2021topologically}. Ordered structures and collective behaviors observed in active matter can only be explained by non-equilibrium statistical physics\citep{caprini2025odd,musacchio2025self,
caprini2025bubble,marconi2024active}. Numerous theoretical models have been applied to explore the complex characteristics of active matter\citep{shaebani2020computational,vicsek1995novel,
toner2005hydrodynamics,cates2013active,romanczuk2012active}. Among these, active Brownian particles (ABPs)---unlike conventional Brownian particles that only move passively---can generate their own propulsion\citep{romanczuk2012active,caprini2020spontaneous,yang2023coherent,fily2012athermal,
redner2013structure,digregorio2018full,elgeti2013wall,fily2014dynamics,
wysocki2020capillary,takatori2014swim}. When the packing fraction and self-propulsion exceed a critical value, ABPs undergo motility-induced phase separation (MIPS) even in the absence of attractive interactions\citep{redner2013structure,buttinoni2013dynamical}. This process results in a phase coexistence of low- and high-density phases---a behavior not observed in equilibrium systems. 

Recent studies reveal that velocities of particles within the high-density phase of ABP systems organize into some aligned or vortex-like velocity-correlation regions\citep{caprini2020spontaneous,yang2023coherent,caprini2020hidden,
caprini2021spatial,marconi2021hydrodynamics}, or even transition to a flocking state\citep{caprini2023flocking}. This finding challenges the earlier assumption that anisotropic interactions between active units are essential for collective motion\citep{ginelli2010large,kudrolli2008swarming,vicsek1995novel}. In addition, several studies reported that the collective motion disappears in thermo-ABP systems as thermal noise increases\citep{caporusso2020motility,negro2022hydrodynamic}. This finding has generated considerable discussion\citep{szamel2021long}. Yang and coworkers have demonstrated that these conflicting results stem from an incorrect treatment in the analysis the motion correlation. While thermal noise contributes a large random component to particles' instantaneous velocities, the inherent motion correlation can be extracted from short-time-averaged velocities, which represent the underlying, real particle motion \citep{yang2023coherent}. They show that collective motion is general, whether thermal noise is present or not, and the velocity-correlation regions are closely related to some kinds of ordered clusters within the solid-like high-density phase. However, the characteristics of these ordered clusters, as well as the mechanisms that induce motion correlation, remain unclear.

In this work, we show that the high-density phase of ABPs is composed of many well-identified, single-orientation crystalline (hexatic) domains. All particles within the crystalline domain move as a coherent unit, exhibiting nearly identical translation or rotation under the net active force or torque acting on the domain. This collective behavior gives rise to the observed motion-correlation region. Particles at the edges of domains have active forces tending to point inward, thus stabilizing and sustaining the domains. The spatial extent of crystalline domains coincides with that of the velocity-correlation regions, and the gyration radius of the domains closely matches the velocity-correlation length, confirming our conclusions.   

This article is structured in the following way. In Section 2, we introduce the simulation details. The main results are shown in Section 3. Finally, a brief conclusion is presented in Section 4. 

\section{Simulation}
Consider a two-dimensional system comprising $N=10,000$ active Brownian particles (ABPs). The stochastic dynamics of ABPs is described by two coupled overdamped Langevin equations\citep{redner2013structure},
\begin{equation}
\label{eq1}
\dot{\bm{r}}_i = -D \beta \bm{\nabla}_i U + D \beta f \bm{n}_i + \sqrt{2 D}\bm{\eta}_i,
\end{equation}

\begin{equation}
\label{eq2}
\dot{\theta}_i=\sqrt{2D_r}\eta^R_i.
\end{equation}
Here, $\dot{\bm{r}}_i$ is the time derivative of the position vector ${\bm{r}}_i$ of the $i$-th particle. $D$ is the translational diffusion constant, $\beta =\frac{1}{k_BT}$ with $k_B$ the Boltzmann constant and $T$ temperature. $\bm{\nabla}_i U$ is gradient of the total interaction potential energy function $U = \sum_{j\ne i} u(r_{ij})$, with the purely repulsive Lennard-Jones potential $u(r_{ij}) = 4\epsilon [(\frac{\sigma}{r_{ij}})^{12}-(\frac{\sigma}{r_{ij}})^{6}]+\epsilon$ while the particle-pair distance $r_{ij} < 2^{\frac{1}{6}}\sigma$, and zero, otherwise. $f$ is the magnitude of active force, and $\bm{n}_i=(cos \theta_i, sin \theta_i)$ is its direction vector. Stochastic force $\bm{\eta}_i$ and torque $\eta^R_i$ are both modeled as zero-mean Gaussian white noise satisfying $\langle\eta^{\mu}(t)\eta^{\nu}(t')\rangle=\delta_{\mu \nu}\delta(t-t')$, and $D_r$ is the rotational diffusion constant, with  $D_r = \frac{3D}{\sigma^2}$ based on the Stokes-Einstein equation. We set $\epsilon$, \ $\sigma$,\ $\tau=\frac{\sigma^2}{D}$ as units of energy, length and time, respectively, and we also choose $k_B T=\epsilon$ in this work. Each simulation trajectory is run for $250 \tau$, and the integrate time step is equal to $10^{-5}\tau$. The segment from $0$ to $150\tau$ is designated as equilibrium run, and all samples are collected within the time interval $[150\tau, 250\tau]$. With the P{\'e}clet number fixed at $P_e=\frac{f \sigma}{k_B T} = 80$ or $100$ and varied area packing fraction $\phi=\frac{N \pi \sigma^2}{4S}$ (where $S$ is the area of the simulation box), the system exhibits different stable states.

\section{Results}

\begin{figure}[ht]
\includegraphics[width=0.8\textwidth]{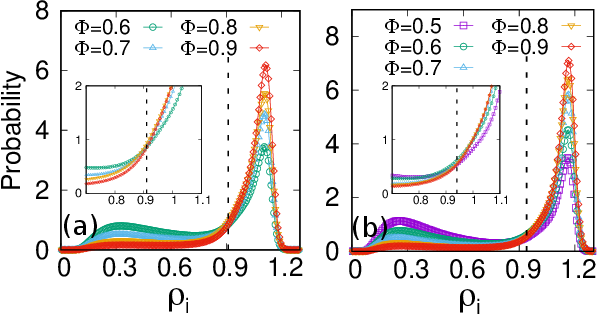}
\caption{\small Probability distribution functions (PDFs) show the local density at different packing fractions for Pe = 80 (left) and Pe = 100 (right). The isosbestic point of these PDFs is indicated by the dashed line. Inset: An enlarged view near the isosbestic point.}
\label{Fig.1}
\end{figure}

We employ the order parameter $\rho_i = \frac{\pi \sigma^2}{4 v_i}$ to characterize the local density of each particle, where the individual cell volume $v_i$ is extracted from the Voronoi tessellation algorithm\citep{rycroft2009voro++}. The resulting probability distribution functions (PDFs) of $\rho_i$ are displayed in Fig.\ref{Fig.1}. All PDFs exhibit a pronounced bimodal shape corresponding to the motility-induced phase separation (MIPS), compatible with the previous results\citep{redner2013structure}. The isosbestic point---the common intersection of the PDFs---serves as a criterion for distinguishing the dense and dilute phases\citep{paolantoni2009tetrahedral}. Accordingly, we set the division at $\rho_i=0.91$ for Pe = 80 and $\rho_i=0.94$ for Pe = 100. 

For analyzing the structure of high-density phases, we calculate the bond-orientational order parameter $\psi_i=\frac{1}{N_i}\sum_j e^{i 6 \theta_{ij}}$\citep{redner2013structure}, where $N_i$ represents the number of nearest neighbors of the particle $i$ (identified by Voronoi tessellation algorithm\citep{rycroft2009voro++}). In this formula, $\theta_{ij}$ signifies the angle between the x-axis and the bond connecting particles $i$ and $j$, with particle $j$ being one of particle $i$'s nearest neighbors. While the absolute value of $\psi_i$ measures the deviation of the neighboring arrangement from the hexatic lattice, its argument indicates the orientation of the lattice\citep{caporusso2020motility,bernard2011two}. 

\begin{figure}[ht]
\includegraphics[width=0.9\textwidth]{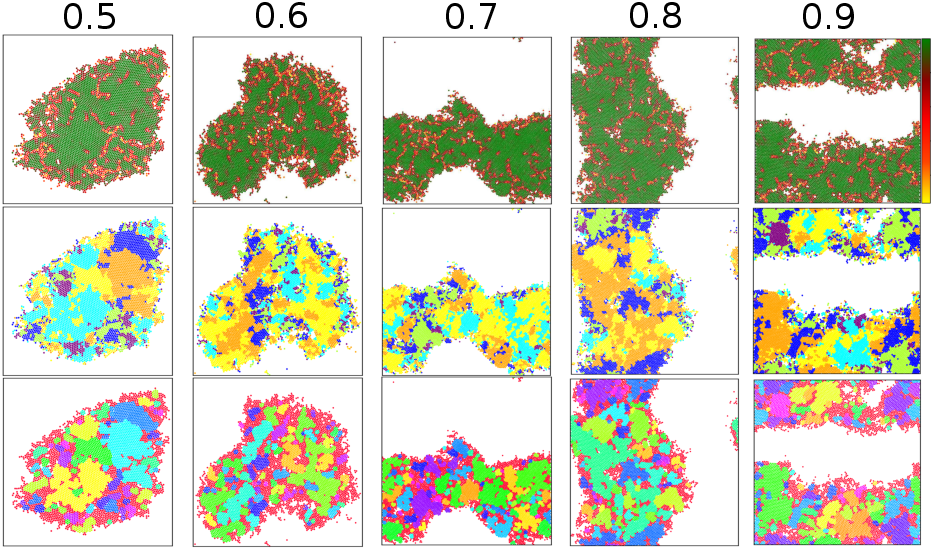}
\caption{\small Structure of the high-density phase at different packing fractions for Pe = 100. In the top row, particles are colored according to the magnitude of the bond-orientational order parameter, $|\psi_i|$. Green particles have a locally hexagonal environment, while the others are defects. In the middle row, particles are colored according to the arguments of $\psi_i$. The bottom row shows the single-orientation crystalline domains, obtained by the DBSCAN clustering algorithm. Red points represent noise particles that do not belong to any domains.}
\label{Fig.2}
\end{figure}  

We color particles inside the high-density phase according to the value of $|\psi_i|$, which approaches $1$ when the particle has a hexatic neighbor environment and is lower for lattice defects. The top row of Fig.\ref{Fig.2} shows that the high-density phase forms a hexatic lattice with some defects. We further assign distinct colors to different parts of the high-density phase based on the argument of $\psi_i$, thus creating clusters with these defects as their boundaries (middle row, Fig.\ref{Fig.2}). Here, the arguments are uniformly divided into $n$ bins, each part with a distinct color. Following the suggestion of Ref.\citep{caporusso2020motility}, we set $n = 6$. 

By using the DBSCAN clustering algorithm\citep{ester1996density}, we identify well-separated domains (bottom row, Fig.\ref{Fig.2}). The colored regions represent the different domains and the red particles surrounding them are classified as noise points that do not belong to any domain. Each domain corresponds to a hexatic lattice with a specific orientation---a single-orientation crystalline domain. A comparison with the top row reveals that these domains are spatially coincident with the hexatic clusters and that the noise points correspond to defects. Here, the clustering algorithm relies on two critical parameters: $eps$ (the maximum distance between two particles considered neighbors) and $min\_samples$ (the minimum number of particles---including the point itself---within the neighborhood of the core point). In this work, we set $eps = 1.3\sigma$, which encloses the first shell of neighbors, and $min\_samples = 7$, signifying that the core point has at least six neighbors in the first shell.   

\begin{figure}[ht]
\includegraphics[width=0.8\textwidth]{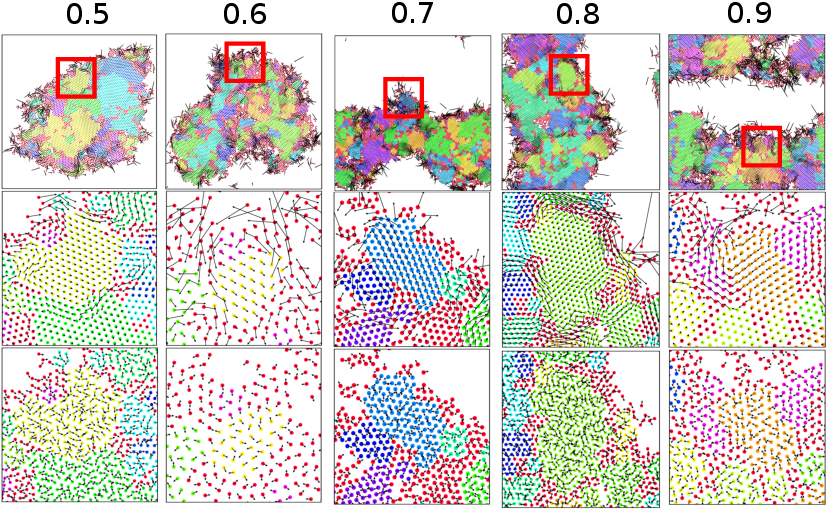}
\caption{\small Structure and collective motion in the high-density phase. The top row shows the (scaled) velocities superimposed on crystalline domains at different packing fractions for Pe = 100. Zoomed views of the areas marked by red squares are shown in the middle row. The bottom row displays the same conformation, with black arrows indicating the direction of the active forces.}
\label{Fig.3}
\end{figure}  

The velocity of particle $i$ at time $t$ is defined as $\bm{V}_i(t, \Delta t)=\frac{\bm{r}_i(t+\Delta t)-\bm{r}_i(t)}{\Delta t}$, where $\bm{r}_i(t)$ is the position vector and $\Delta t$ is the lag time. As recommended by Ref.\citep{yang2023coherent}, the lag time $\Delta t$ should be chosen to maximize the velocity-correlation order parameter $Q= \langle 1-2 \sum_{ij} \frac{d_{ij}}{N_i \pi} \rangle_h$ \citep{caprini2020spontaneous}. Here, the sum runs over the $N_i$ nearest neighbors of particle $i$, and $d_{ij}$ is the angle between the velocities of particle $i$ and that of its neighbors. The average $\langle \dots \rangle_h$ is taken over all particles within the high-density phase. In our simulations, the order parameter $Q$ reaches its maximum around $\Delta t=0.1 \tau$ for all the packing fractions, consistent with the findings reported in Ref.\citep{yang2023coherent}. We therefore use this lag time for computing all the velocities.

The top row of Fig.\ref{Fig.3} shows particles' velocities superimposed on crystalline domains. The velocities inside each domain exhibit coherent vortex-like or aligned pattern. Conversely, defect particles (i.e.noise points) located between domains display disordered velocities, with magnitudes substantially larger than those within domains. The middle row provides a magnified view of the red squares in the top row, showing an individual domain and its surroundings. It is evident that the inner particles move coherently to form a perfect velocity-correlation region, whereas edge particles exhibit abrupt velocity changes. In the bottom row, the directions of active forces are overlaid on domains. Unlike the velocity field, there is no clear relationship between active forces. A detailed analysis of force correlations will be given later.

\begin{figure}[ht]
\includegraphics[width=0.9\textwidth]{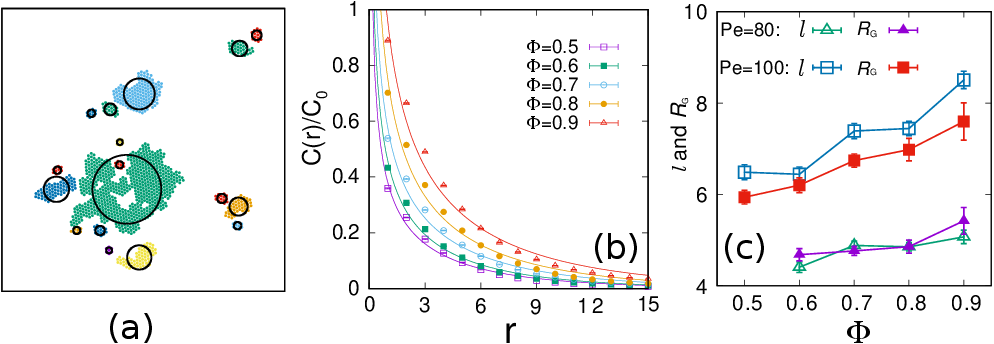}
\caption{\small Size of structure-ordered domains and spatial correlation of motion. (a) Single-orientation crystalline domains in one part of the high-density phase at $\phi=0.5$ and Pe = 100 (noise points omitted). Each black circle is positioned at the centroid of a domain, with its radius equal to the gyration radius of the domain. (b) Spatial velocity correlations in the high-density phase at different packing fractions for Pe = 100. The solid line is a theoretical fitting (see main text) to the data. To make the correlations easier to distinguish, each correlation is divided by a constant $c_0$ (which does not affect the correlation length). (c) Comparison between the velocity-correlation length and the weighted-average gyration radius for Pe = 80 and Pe = 100.}
\label{Fig.4}
\end{figure}  

We continue by comparing the size of crystalline domains and the velocity-correlation length. Domains identified in one part of the high-density phase are presented in Fig.\ref{Fig.4}(a). Each black circle is positioned at the centroid of the domain, with a radius equal to the domain's gyration radius. The gyration radius for the $p$-th domain is calculated as $R^p_G=\sqrt{ N_p^{-1} \sum_{q \in p} (\bm{r}_q-\bm{r}_p^{c})^2}$, here $N_p$ is the number of particles within the $p$-th domain, $\bm{r}_q$ is a particle's position vector, and $\bm{r}_p^{c}$ represents the position vector of the domain's centroid. We calculate the spatial velocity correlation for the high-density phase using $C(r)=\langle \bm{V}(0) \cdot \bm{V}(r)\rangle_h$. These results are displayed in Fig.\ref{Fig.4}(b). The solid line represents a fit to the data with the function $\frac{A}{r^{1/2}}e^{-r/l}$\citep{caprini2020hidden,caprini2021spatial}, where $A$ is a fitting parameter and $l$ is the correlation length. Fig.\ref{Fig.4}(c) shows a comparison between the velocity-correlation length and the weighted-average gyration radius, $R_G=\frac{\sum_p N_p R_G^p}{\sum_p N_p}$. The velocity-correlation length closely matches the gyration radius across all packing fractions. This agreement strongly supports that the spatial extent of the crystalline domains coincides with the velocity-correlation regions. 

\begin{figure}[ht]
\includegraphics[width=0.8\textwidth]{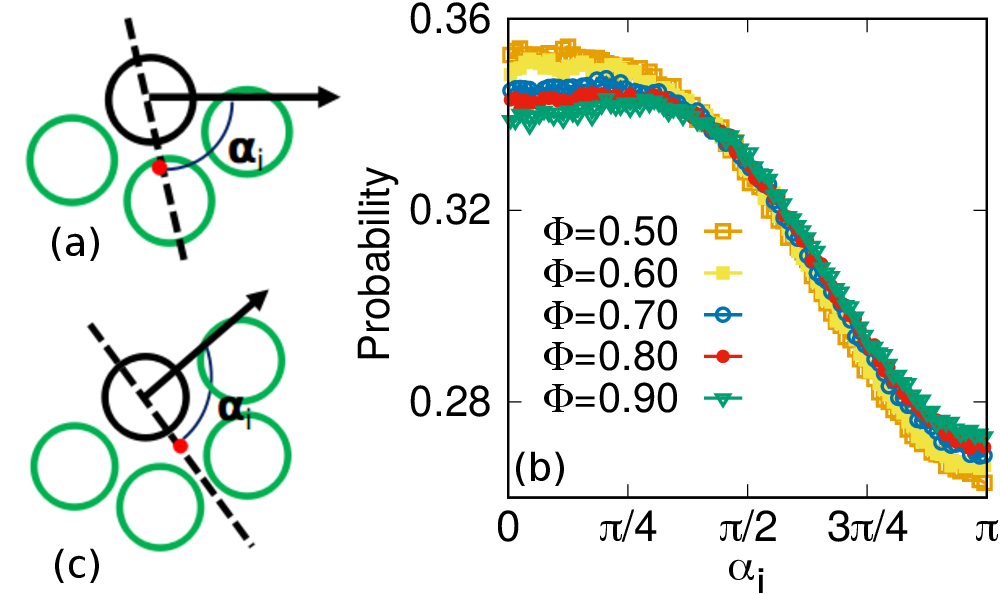}
\caption{\small Direction of active force at boundary of domains. (a) Schematic defining the angle $\alpha_i$. The particle at the domain edge ($i$, black circle) and its neighbors (green circles) are shown. The red point marks the centroid of these neighbors. (b) Probability density functions of $\alpha_i$ at various packing fractions for Pe = 100. (c) A particle trapped at the edge of a domain requires a larger escape angle for larger domains.}
\label{Fig.5}
\end{figure}  

To investigate the formation of crystalline domains, we compute the angle $\alpha_i$ following the method of Ref.\citep{yang2023coherent}. The geometry of this parameter is illustrated in Fig.\ref{Fig.5}(a). The black circle represents an edge particle $i$ of a domain; the green circles denote its nearest neighbors. The black arrow shows the direction of the active force on this particle, and the red dot indicates the center of mass (CM) of all its nearest neighbors. We define $\alpha_i$ as the angle between the active force vector and the bond connecting particle $i$ to the CM. When $\alpha_i$ is acute, particle $i$ tends to compress the domain; otherwise, it tends to leave from the domain. Fig.\ref{Fig.5}(b) shows the probability density functions of $\alpha_i$ for different packing fractions. Notably, all distributions peak at acute angles, indicating that the active forces on edge particles are predominantly point inward. This inward force compacts the domain, resulting in its solid-like properties. As the packing fraction increases, the distribution of $\alpha_i$ broadens slightly. This broadening is mainly attributed to the growth in domain size [see the gyration radius in Fig.\ref{Fig.4}(c)]. Larger domains typically provide each edge particle with more neighbors. A comparison of Fig.\ref{Fig.5}(a) and (c) reveals that edge particles with more neighbors require a larger $\alpha_i$ to escape from the domain.

\begin{figure}[ht]
\includegraphics[width=0.9\textwidth]{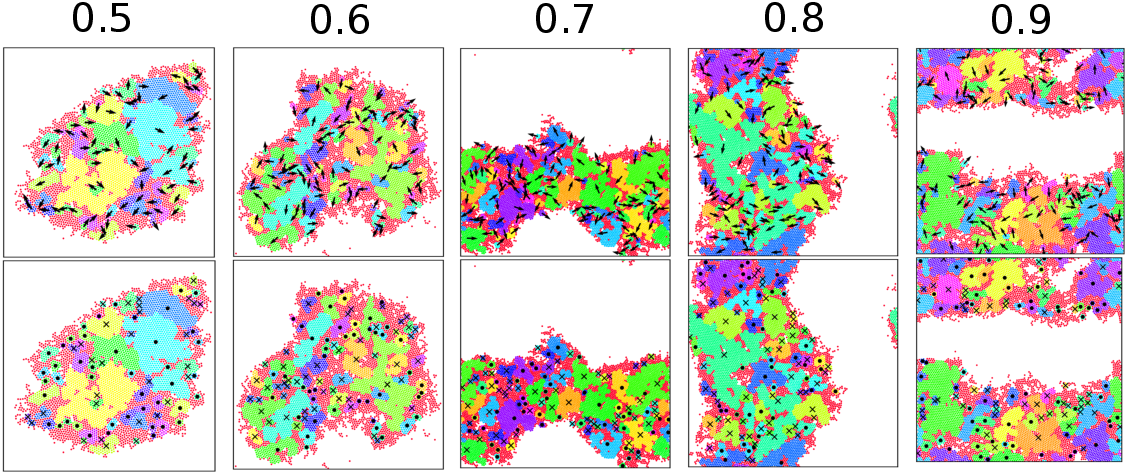}
\caption{\small Crystalline domains at different packing fractions for Pe = 100. Arrows indicate the direction of their centroid velocities (top row). In the bottom row, black dots and crosses mark domains with positive and negative angular velocity, respectively.}
\label{Fig.6}
\end{figure}  

To study the motion of each crystalline domain as a single entity, we calculate its centroid velocity and angular velocity. For the $p$-th domain, the centroid velocity is given by $\bm{V}_p^c=N_p^{-1}\sum_{q\in p} \bm{V}_q$, where $N_p$ is the number of particles within the domain and $\bm{V}_q$ is the velocity of particle $q$. Its angular velocity is computed as: $\bm{\omega}_p =\frac{\sum_{q\in p} (\bm{r}_q-\bm{r}_p^c)\times \bm{V}_q}{\sum_{q\in p} (\bm{r}_q-\bm{r}_p^c)^2} = \frac{\sum_{q\in p} (\bm{r}_q-\bm{r}_p^c)\times \bm{V}_q}{k_p}$, where $\bm{r}_q$ is the position vector of particle $q$, $\bm{r}_p^c$ is the centroid, and $k_p=\sum_{q\in p} (\bm{r}_q-\bm{r}_p^c)^2$. In the top row of Fig.\ref{Fig.6}, black arrows indicate the direction of the centroid velocity. In the bottom row, domains with positive angular velocity are marked with black dots, while those with negative angular velocity are marked with crosses. 

\begin{figure}[ht]
\includegraphics[width=0.8\textwidth]{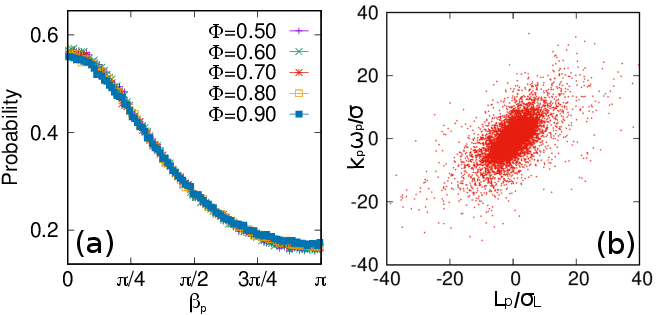}
\caption{\small  Relation between motion of crystalline domains and the net active force and torque acting on them. (a) Probability density functions of $\beta_p$ at various packing fractions for Pe = 100. (b) Relationship between the weighted angular velocity of domains and the net active torque acting on them at $\phi=0.5$ and Pe = 100. The symbols $\sigma$ and $\sigma_L$ denote the standard deviations of the weighted angular velocity and torque, respectively.}
\label{Fig.7}
\end{figure}

Then, we explore the cause of the translation and rotation of crystalline domains. The net active force is defined as $\bm{F}_p=\sum_{q\in p} \bm{f}_q$, where $\bm{f}_q$ is the active force acting on particle $q$ within $p$-th domain. The align angle between the centroid velocity $\bm{V}_p^c$ and the net active force $\bm{F}_p$ is denoted as $\beta_p$. The probability density function of $\beta_p$ is shown in Fig.\ref{Fig.7}(a). All distributions peak at acute angles, indicating that each domain moves along the direction of the net active force; in other words, the domain is propelled by the net active force. The relationship between the weighted angular velocity, $k_p \bm{\omega}_p$, and the net active torque is shown in Fig.\ref{Fig.7}(b). Here, the net active torque is calculated as $\bm{L}_p=\sum_{q\in p} (\bm{r}_q-\bm{r}_p^c)\times \bm{f}_q$. A proportional relation is observed, indicating that the rotation of the domain is controlled by the net active torque. 

\section{Conclusion}
In summary, we show that the solid-like high-density phase in the active Brownian particle system is composed of multiple single-orientation crystalline domains. The boundary particles of these crystalline domains have inward-oriented active forces to maintain these structure-ordered domains. Each domain then translates or rotates collectively as a rigid body under the net active force or torque. The defective particles that separate various domains allow the domains to shift relative to one another, thereby breaking the inter-domain motion correlation. Our findings offer a perspective for understanding motion correlation in active matter through its ordered structure. This perspective suggests the possibility of directing collective motion by inducing ordered structures, rather than relying solely on the anisotropic properties of the particles. 

\section{Acknowledgments}
We thank Mingcheng Yang for helpful suggestions and this work is supported by the Initial Scientific Research Fund and the Innovation Team Fund of Mianyang Teachers' College (Grants No. QD2020A03 and No. CXTD2023PY01).

\bibliographystyle{apsrev4-1}
\bibliography{ref.bib} 

%%%%%%%%%%%%%%%%%%%%%%%%%%%%%%%%%%%%%%%%%%%%%%%%%%%%%%%%%%%%%%%%%%%%%%%%%%%%%%%%%%%%%%%%%%%%%%%%%

\end{document}